\documentstyle[12pt]{article}

\newcommand{\rf}[1]{(\ref{#1})}
\newcommand{\bea}{\begin{eqnarray}}
\newcommand{\eea}{\end{eqnarray}}

\newcommand{\e}{{\rm e}}
\renewcommand{\d}{{\rm d}}

\newcommand{\m}{\mu}

\newcommand{\ep}{\varepsilon}

\newcommand{\del}{\delta}
\newcommand{\Del}{\Delta}

\newcommand{\oh}{\frac{1}{2}}

\newcommand{\tr}{{\rm Tr}\;}
\newcommand{\ra}{\right\rangle}
\newcommand{\la}{\left\langle}
\newcommand{\prt}{\partial}

\newcommand{\cD}{{\cal D}}

\newcommand{\tdqg}{two-dimensional quantum gravity}

\def\void{}
\def\labelmark{}

\newenvironment{formula}[1]{\def\labelname{#1}
\ifx\void\labelname\def\junk{\begin{displaymath}}
\else\def\junk{\begin{equation}\label{\labelname}}\fi\junk}%
{\ifx\void\labelname\def\junk{\end{displaymath}}
\else\def\junk{\end{equation}}\fi\junk\labelmark\def\labelname{}}

{\ifx\void\labelname\def\junk{\end{array}\end{displaymath}}
\else\def\junk{\end{array}\right.\end{equation}}
\fi\junk\labelmark\def\labelname{}\def\junk{}
}

\newcommand{\beq}{\begin{formula}}
\newcommand{\eeq}{\end{formula}}
\newcommand{\beqv}{\begin{formula}{}}

\begin{document}
\topmargin 0pt
\oddsidemargin 5mm
\headheight 0pt
\headsep 0pt
\topskip 9mm

\hfill   NBI-HE-97-62

\hfill   TIT/HEP-381

\hfill   December 1997

\begin{center}
\vspace{24pt}
{ \large \bf The Spectral Dimension of 2D Quantum Gravity}

\vspace{24pt}

{\sl Jan Ambj\o rn}$\,^{a,}$\footnote{email ambjorn@nbi.dk},
{\sl Dimitrij Boulatov}$\,^{b,}$\footnote{email boulatov@heron.itep.ru},
{\sl Jakob L.\ Nielsen}$\,^{a,}$\footnote{email jlnielsen@nbi.dk }\\
{\sl Juri Rolf}$\,^{a,}$\footnote{email rolf@nbi.dk} and 
{\sl Yoshiyuki Watabiki}$\,^{c,}$\footnote{email 
watabiki@th.phys.titech.ac.jp}  

\vspace{24pt}

$^a$~The Niels Bohr Institute, \\
Blegdamsvej 17, \\
DK-2100 Copenhagen \O , Denmark, 

\vspace{24pt}
$^b$~Inst. Theor. Exp. Phys., \\
    B. Cheremushkinskaya ul. 25, \\
    117259 Moscow, Russia, 

\vspace{24pt}

$^c$~Department of Physics, \\
Tokyo Institute of Technology, \\ 
{Oh}-okayama, Meguro, Tokyo 152, Japan

\end{center}

\vfill

\begin{center}
{\bf Abstract}
\end{center}

\vspace{12pt}

\noindent
We show that the spectral dimension $d_s$  of two-dimensional 
quantum gravity coupled to Gaussian fields is two for all values of 
the central charge $c \leq 1$. 

The same arguments provide a simple proof of the 
known result $d_s= 4/3$ for branched polymers.

\vfill

\newpage

\section{Introduction}

One of the main objectives of two-dimensional quantum gravity 
is to gain an understanding of the fractal structure of space-time.
The fractal structure reflects true quantum phenomena. 
In the case of pure \tdqg, i.e.\ when the central charge $c$ of conformal 
matter coupled to gravity is equal to zero, we know that the {\it intrinsic} 
Hausdorff dimension is four (not two as for smooth, fixed geometries)
\cite{kkmw,aw}, and for
conformal matter with $c= -2$ it is $3.56\cdots$ \cite{kksw,many1,many2}. 
These results can be obtained by calculating
the average volume $V(R)$ of balls of geodesic radius $R$. However, there
are other definitions of the fractal structure of space-time which might
reflect other aspects of the average space-time one encounters in
two-dimensional quantum gravity. One such measure of the fractal 
properties of space-time is the so-called {\it spectral} dimension,
which can be defined in an reparameterization invariant way and 
consequently makes sense in quantum gravity.

The most intuitive definition of the spectral dimension 
is based on the diffusion equation on a (compact) manifold with 
metric $g_{ab}$. Let $\Del_g$ denote  the Laplace-Beltrami 
operator corresponding to $g_{ab}$. The probability distribution
$K(\xi,\xi';T)$ of diffusion is related to the massless scalar propagator 
$(-\Del_g)^{-1}$ by \footnote{Since we consider compact manifolds 
the Laplace-Beltrami operator $\Del_g$ has zero modes. Eq.\ \rf{1.1}
should be understood with these zero modes projected out. This is indicated 
with a prime.}
\beq{1.1}
\la \xi'\Big| (-\Del_g)^{-1} \Big| \xi\ra' 
= \int_0^\infty\! dT \; K'(\xi,\xi';T) .
\eeq
In particular, the average return probability distribution at time $T$ 
has the following small $T$ behavior:
\beq{1.2}
RP'_g(T) \equiv \frac{1}{V_g}\int\! \d^d \xi \sqrt{g} \; K'(\xi,\xi;T) \sim 
\frac{1}{T^{d/2}}(1+O(T)),
\eeq
where $V_g$ denotes the volume of the compact manifold with metric $g$.
The important point, in relation to quantum gravity, is that $RP'_g(T)$ 
is invariant under reparameterization. Thus the quantum average over
geometries can be defined:
\beq{1.2a}
RP'_V(T) \equiv \frac{1}{Z_V}  \int\! \cD [g]_V\;\e^{-S_{\rm eff}([g])} 
RP'_g(T),
\eeq
where $Z_V$ denotes the partition function of quantum gravity for
fixed space-time volume $V$ (see \rf{2.1} for more details about 
$Z_V$ and $\cD [g]_V$), and $S_{\rm eff}([g])$ denotes the effective
action of quantum gravity after the integration over possible matter
fields. The {\it spectral}  dimension $d_s$ in quantum gravity is 
now defined by the small $T$ behavior of the functional average $RP'_V(T)$ 
\beq{1.2b}
RP'_V(T) \sim \frac{1}{T^{d_s/2}}(1+ O(T)).
\eeq
The $O(T)$ term in \rf{1.2} has a well known asymptotic expansion 
in powers of $T$, where the coefficient of $T^r$ is an integral 
over certain powers and certain contractions of the curvature tensor.
This asymptotic expansion breaks down when $T \sim V^{2/d}$ at which 
point the exponential decay in $T$ of the heat kernel $K$ takes over.
If we average over all geometries  as in \rf{1.2a}, it is natural to 
expect that the only invariant left will be the volume $V$ which is kept
fixed. Thus we expect that we can write
\beq{1.3}
RP'_V(T) = \frac{1}{T^{d_s/2}} F\Big(\frac{T}{V^{2/d_s}}\Big),
\eeq  
where $F(0) > 0$ and $F(x)$ falls off exponentially fast for
$x \to \infty$. 

For fixed manifold of dimension $d$ and a given smooth geometry $[g]$ 
we have $d=d_s$ by definition. The functional average can {\it a priori} 
change this, i.e.\ the dimension of $T$ can become anomalous. 
A well known example of similar nature can be found 
for the ordinary free particle. In the path integral representation
of the free particle, any smooth path of course has fractal dimension
equal to one. Nevertheless the short distance properties of the free 
particle reflects that the generic path contributing to the path integral 
has fractal dimension (the {\it extrinsic} 
Hausdorff dimension -- in the target space $R^D$) 
$D_H=2$ with probability one. In the same way the functional integral over 
geometries might change $d_s$ from the ``naive'' value $d$.
In \tdqg\ it is known, as mentioned above, 
that the intrinsic Hausdorff dimension {\it is} different from $d=2$
and the generic geometry is in this sense fractal, with probability one.
When one considers diffusion on fixed fractal structures 
(often embedded in $R^D$) it is well known 
that $d_s$ can be different from both $D$ and the fractal dimension $d_h$
of the structure. If $\del$ denotes the so-called anomalous gap exponent,
defined by the relation between the diffusion time $T$ and 
the average spread of diffusion on the fractal structure, but measured in 
$R^D$:
\beq{1.2c}
\la r^2(T)\ra \sim T^{2/\del},
\eeq
then the relation between the fractal dimension ({\em intrinsic}
Hausdorff dimension) $d_h$ of the structure,
the spectral dimension of the diffusion and the gap exponent is:
\beq{1.2d}
d_s = \frac{2d_h}{\del}.
\eeq
If $\del$ is not anomalous, i.e.\ $\del =2$ as for diffusion on a smooth 
geometry, we have $d_s = d_h$, which is the analogue of $d_s =d$ for 
fixed smooth geometries. However, in general $\del \neq 2$ (for a review
of diffusion on fractal structure, see e.g.\ \cite{review}).
    
In principle the above definitions apply for any theory of 
quantum gravity (see for instance \cite{aj,book1,book2} for definitions in the 
case of four dimensional simplicial quantum gravity).
However, only for \tdqg\ we have presently a well defined theory
which allows us to make detailed calculations. In the following 
we will not consider higher dimensional quantum gravity.

\section{Spectral dimension for 2d quantum gravity}

We will derive a simple relation between the spectral dimension 
and the extrinsic Hausdorff dimension for dynamical self-similar 
systems like \tdqg,  branched polymers etc. Since the extrinsic 
Hausdorff dimension is known for these systems it will allow a
determination of the spectral dimension. As a typical example 
of such models (and maybe the most interesting), we consider \tdqg.
The partition function for \tdqg\ coupled to $D$ Gaussian fields
$X_\m$ is given by 
\beq{2.1}
Z_V = \int\! \cD [g]_V \cD [X_\m]_{\rm cm}\, 
\e^{-\!\int\! \d^2 \xi \sqrt{g} \, g^{ab} \prt_a X_\m \prt_b X_\m},
\eeq
where $\int \cD [g]_V$ denotes the integration 
over {\it geometries}, i.e.\ equivalence classes of metrics on 
the two-dimensional manifold of fixed space-time volume $V$ 
of the manifold, while $\cD [X_\m]_{\rm cm}$ denotes the functional 
integration over the $D$ Gaussian fields $X_\m$, but with the center of 
mass fixed (to zero). The extrinsic Hausdorff 
dimension $D_H$ is usually defined as 
\beq{2.2}
\la X^2 \ra_V \sim V^{2/D_H}~~~~{\rm for}~~~V \to \infty,
\eeq
where
\beq{2.3}
\la X^2 \ra_V \equiv \frac{1}{Z_V} \int\! \cD [g]_V \cD [X_\m]_{\rm cm}\,
\e^{-\!\int\! \d^2 \xi \sqrt{g} \, g^{ab} \prt_a X_\m \prt_b X_\m}\;
\frac{1}{DV}  \int\! \d^2 \xi \sqrt{g} \, X_\m^2(\xi) .
\eeq
The Gaussian action in $X$ implies that:
\bea
\label{2.4}
\la X^2 \ra_V &=& \frac{1}{DV Z_V} \frac{\partial}{\partial \omega}
\int\! \cD [g]_V \cD [X_\m]_{\rm cm}\,\e^{-\!\int\! \d^2 \xi \sqrt{g}
  \, g^{ab} \prt_a X_\m \prt_b X_\m + \omega \!\int\! \d^2 \xi \sqrt{g}
  X_\m^2(\xi)}\Bigg\vert_{\,\omega=0} \nonumber\\
&=& \frac{1}{DV Z_V} \frac{\partial}{\partial \omega}
\int\! \cD [g]_V\Big({\rm det}' (-\Del_g-\omega)\Big)^{-\!D/2}\;
\Bigg\vert_{\,\omega=0} \nonumber\\
&=& \frac{1}{2 V Z_V} \int\! \cD[g]_V
\Big({\rm det}' (-\Del_g)\Big)^{-\!D/2}\;
\tr'\! \left[\frac{1}{-\Del_g}\right]\nonumber\\
& =& 
\frac{1}{2V}\la\tr'\!\left[\frac{1}{-\Del_g}\right]\ra_V
\eea
where the primes on the determinants and traces again mean that zero modes are
excluded. Formula \rf{2.4} is used to define $\la X^2 \ra_V$ when $D$ is 
non-integer.

Using \rf{1.1} and \rf{1.3} we get 
\beq{2.5}
\la X^2 \ra_V = \oh \int_0^\infty\! dT \; \frac{1}{T^{d_s/2}} 
F\Big(\frac{T}{V^{2/d_s}}\Big) \sim V^{2/d_s -1},
\eeq
for $V$ going to infinity. From \rf{2.2}
we now conclude that 
\beq{2.7}
\frac{1}{d_s} = \frac{1}{D_H} +\frac{1}{2}.
\eeq

Several remarks are in order. Strictly speaking the above derivation 
assumes that $d_s < 2$. If $d_s > 2$ we have to introduce a
small-$T$ cut-off $\ep$ in \rf{2.5}. In this case it is convenient to 
consider instead $\la (X^2)^n \ra_V \sim V^{2n/D_H}$ where 
$n = [d_s/2]+1$ for non-integer $d_s$. It is then easy to show 
that the leading large $V$ 
behavior on the right hand side of the equation corresponding to 
\rf{2.5} will be $V^{2n/d_s-1}$ and we get 
\beq{2.7a}
\frac{1}{d_s} = \frac{1}{D_H} +\frac{1}{2n} 
~~~~{\rm for}~~~ 2n-2<d_s<2n 
~~\hbox{( $n=1,2,\ldots$ )} .
\eeq
In the study of diffusion on fixed fractal structures one 
usually encounters $d_s <2$. In the following we will 
{\it assume} $d_s \leq 2$ due to the following reasoning. 
We  expect that $d_s \to 2$ for $D \to -\infty$
since large negative $D$ implies that a saddle point calculation
of \rf{2.3} around a fixed geometry should be reliable. A strictly 
fixed geometry implies $d_s =2$ and $D_H = \infty$ (in agreement with 
\rf{2.7}). Also the saddle point calculation results in $D_H=\infty$ and 
should be valid in a neighborhood of $D=-\infty$. Hence $d_s =2$ in a 
neighborhood of $D=-\infty$. 
If doing anything, one would expect fluctuating geometries 
to decrease $D_H$ since there are many ``degenerate'' geometries where 
$D_H < \infty$, e.g.\ the branched polymer-like geometries to be discussed
later. Thus it is reasonable to assume $d_s \leq 2$.

Once this assumption is made, it follows immediately that 
$d_s = 2$ for all $D \leq 1$ since it is known from Liouville 
theory that $D_H = \infty$. Let us just recall the 
argument\footnote{The treatment in \cite{kawai1} is based on a 
more general scaling assumption than is needed in \tdqg\ and this leaves
open the possibility of a scaling different from the one given here. In 
\cite{kawai2} the treatment was narrowed down to the one presented here.}
\cite{kawai1,kawai2}.
Define the two-point function in random surface theory by:
\beq{2.8}
G(p) = \la \int\!\! \int\! \d^2\xi_1\sqrt{g(\xi_1)} 
\,\d^2 \xi_2 \sqrt{g(\xi_2)} \, \e^{i p (X(\xi_1)-X(\xi_2))}\ra_V 
\eeq
If we use the following definition of $\la X^2 \ra_V$ (which is equivalent to 
\rf{2.3} for large $V$)
\beq{2.9}
\la X^2 \ra_V  = \frac{1}{D V^2} 
\la  \int\!\! \int\! \d^2\xi_1\sqrt{g(\xi_1)}\, \d^2 \xi_2 \sqrt{g(\xi_2)}  
 \Big(X(\xi_1)-X(\xi_2)\Big)^2 \ra_V,
\eeq
it follows that 
\beq{2.10}
\la X^2 \ra_V = -\frac{1}{D V^2} \frac{\prt^2}{\prt p^2} \, 
G(p) \Big|_{p=0}.
\eeq
Since it is known that $G(p)$ behaves as $V^{2-\Del_0(p)}$ in flat 
space with $\Del_0 (p) \propto p^2$, the KPZ formula allows us 
to calculate $\Del(p)$ after coupling to gravity:
\beq{2.11}
\Del(p) = \frac{\sqrt{1-D +24\Del_0(p)}-\sqrt{1-D}}{\sqrt{25-D}-\sqrt{1-D}}.
\eeq
It follows that for $D < 1$ we have 
\beq{2.12}  
\la X^2 \ra_V \sim \log V,
\eeq
while for $D =1$
\beq{2.13}
\la X^2 \ra_V \sim \log^2 V.
\eeq
In both cases $D_H = \infty$ and thus $d_s=2$ from \rf{2.7}.\\

\section{Discussion}

We have shown under mild assumptions 
that the spectral dimension $d_s=2$ for \tdqg\ coupled to $D$ 
Gaussian fields. If we assume that the 
spectral dimension is a function only of the central charge of the 
matter fields coupled to \tdqg, it follows that $d_s$ is always two.
This assumption is corroborated by numerical simulations for pure 
gravity, the Ising model ($c$ = 1/2) coupled to gravity  and the 
three-states Potts model ($c$ = 4/5) coupled to gravity \cite{ajw},
as well as high-statistics simulations for $c=-2$ \cite{many3}.
{}From the point of view of fractal structures the situation is 
most remarkable: the generic manifold is highly fractal when 
defined in the conventional way, using volume $V(R)$ versus 
geodesic distance $R$ as a measure of the fractal dimension for small $R$.
Accordingly the gap exponent $\del$ for diffusion 
is large and anomalous compared to the generic value $\del =2$ 
for a smooth manifold. However,
$\del$ is exactly equal to the anomalus fractal dimension $d_h$ and in this 
way the spectral dimension of the generic, fractal geometry of 
the two-dimensional manifold which appears in the functional integral
is {\it two}, the same as that of a smooth, compact two-dimensional 
manifold!

For $D > 1$ it is generally believed that the two-dimensional surfaces 
degenerate to branched polymers. The Gaussian fields represent an
embedding of these branched polymers into $R^D$ and it is well known 
that $D_H =4$ for the generic branched polymers. We thus conclude 
that the spectral dimension of branched polymers is equal to 4/3, the famous 
Alexander-Orbach  value. The value $d_s=4/3$ as well as formula 
\rf{2.7} was derived for branched polymers by a different 
method in \cite{cates} (see \cite{jw} for a recent more elaborate and
complete proof that $d_s=4/3$). Furthermore it is well known 
that there exists a well defined 
class of {\it multicritical branched polymers} \cite{adj}, very similar to 
the multicritical matrix models. They have $D_H = 2m/(m-1)$, $m=2,3,\ldots$,
where the $m=2$ class contains the ordinary branched polymers 
with positive branching ratios.
For these we obtain $d_s = 2m/(2m-1)$. It is seen that $d_s \to 1$ for
$m \to \infty$. This is in agreement with the fact that these multicritical 
branched polymers approach ordinary random walks for $m \to \infty$.
Clearly, for an ordinary random walk where $D_H =2$, we get $d_s =1$
from \rf{2.7}, as expected. It is also in agreement with a very recent 
more elaborate analysis \cite{jw1}.

Finally we would like to discuss a subtlety buried in the definition 
\rf{1.2a} of $RP'_V(T)$. We have defined the  return probability 
by first calculating it for a fixed geometry and then performing 
the functional integral over geometries. Alternatively, one could have 
used $K_g(\xi,\xi;T)$ to define diffusion as a function of geodesic distance
$R$ by 
\beq{5.1}
K'_V(R;T) = \frac{1}{V Z_V}\int\! \cD [g]_V\e^{-S_{\rm eff}([g])} 
\int \!\d \xi \sqrt{g} \!\int 
\!\d \xi' \sqrt{g'} 
\; \del(d_g(\xi,\xi')-R) K'_g(\xi,\xi';T),
\eeq
where $d_g(\xi,\xi')$ denotes the geodesic distance from $\xi$ to $\xi'$
in the geometry defined by the metric $g$. One would be tempted
to say that $RP'_V(T) = K'_V(0;T)$. However, it is not known whether 
the limit $R \to 0$ commutes with the functional integration.
For the so-called two-point function, which is obtained from 
\rf{5.1} by substituting $1$ instead of $K_g$, it is known that the 
functional interal does {\it not} commute with limit $R \to 0$.
If the limits do not commute we have two inequivalent definitions of
$RP'_V(T)$. Our derivation is just based on the 
assumption that the expression we use satisfies \rf{1.3}.
It is an interesting unsolved problem to understand if the functional 
integration commutes with the $R \to 0$ limit, and which definition of  
$RP'_V(T)$ is correct in case functional integration  and $R\to 0$ are
non-commuting.

\end{document}